\documentclass[onecolumn,prd,aps,amsmath,amssymb]{revtex4}

\usepackage{graphicx}
\usepackage{subfigure}
\usepackage{color}
\usepackage{dcolumn}
\usepackage{bm}

\begin{document}

\title{Effective models of two-flavor QCD: from small towards large $m_q$}
\author{T.~K\"ah\"ar\"a}
\email{topi.kahara@jyu.fi}
\author{K.~Tuominen}
\email{kimmo.tuominen@jyu.fi}
\affiliation{Department of Physics, University of Jyv\"askyl\"a, P.O.Box 35, FIN-40014 Jyv\"askyl\"a, Finland \\
and 
Helsinki Institute of Physics, P.O.Box 64, FIN-00014 University of Helsinki, Finland\\}
\begin{abstract}
We study effective models of chiral fields and Polyakov loop expected to describe the dynamics responsible for the phase structure of two-flavor QCD. We consider chiral sector described either using linear sigma model or Nambu-Jona-Lasinio model and study how these models, on the mean-field level when coupled with the Polyakov loop, behave as a function of increasing bare quark (or pion) mass. We find qualitatively similar behaviors for the cases of linear sigma model and Nambu-Jona-Lasinio model and, relating to existing lattice data, show that one cannot conclusively decide which or the two approximate symmetries drives the phase transitions near the physical point. 
\end{abstract}

\maketitle

\section{Introduction}

It has been proposed that one can understand the interplay between chiral symmetry restoration and center symmetry breaking and their implications on the phase transitions of QCD at finite temperature and density by treating chiral symmetry as almost exact, and center symmetry breaking as a consequence of chiral symmetry restoration and interactions \cite{Mocsy:2003qw}. This picture allows one to explain, in very transparent terms, why deconfinement and chiral symmetry restoration coincide in the theory with fundamental quarks \cite{Kaczmarek:1999mm} while for adjoint quarks the two phase transitions are widely separated \cite{Karsch:1998qj}. For adjoint quarks at finite chemical potential this model framework also predicts interesting multicritical phenomena, see \cite{Sannino:2004ix}. For QCD the standard viewpoint, supported by the physical spectrum, is that in the two-flavor case the chiral symmetry is almost exact whereas the center symmetry is badly broken. However, one might regard this picture too simple, since it does not take into account how rapidly the system departs from the limit of exact chiral symmetry. Consider cranking pion mass up from its physical value. There are basically two options: First, it could be that even for modest masses chiral symmetry remains a good approximation while the center symmetry becomes an approximate symmetry only for a very heavy pion. Alternatively, the second option is that the system departs from the chiral limit very fast, and chiral symmetry becomes very quickly a poor symmetry simultaneously allowing center symmetry to restore rapidly as a function of pion mass. This latter viewpoint has been advocated for in e.g. \cite{Dumitru:2003cf}. A new ingredient of our analysis is the treatment of chiral and Polyakov degrees of freedom within the same model setup simultaneously. Furthermore, in the spirit of our earlier work \cite{Kahara:2008yg}, we consider side-by-side two different realizations of the chiral sector. This will allow us to gain more insight into what extent the quantitative results of these models can be trusted.

We consider linear sigma model and Nambu-Jona-Lasio model \cite{Nambu:1961tp} both augmented with the Polyakov loop. There has been a large body of literature devoted on these models recently \cite{Fukushima:2003fw,Kahara:2008yg,Ratti,Rossner:2007ik,Ciminale:2007ei,Costa:2008yh,Sakai:2009dv}. These models provide a simple and concrete realization for the intertwining of deconfinement and chiral symmetry restoration as explained in general terms in \cite{Mocsy:2003qw}. As discussed in detail in \cite{Kahara:2008yg}, especially at finite densities the numerical output is quite sensitive on which realization of the chiral dynamics is used. Therefore the quantitative results should, when used to draw concusions about QCD dynamics, be interpreted with some care.

In this paper, we continue the investigation started in \cite{Kahara:2008yg}. Here our aim is to study how these models respond when the amount of explicit chiral symmetry breaking is increased. 
For the case of Polyakov-linear sigma model (PLSM), we basically confirm the picture established in \cite{Dumitru:2003cf} and we find that Polyakov-Nambu-Jona-Lasinio (PNJL) model leads to qualitatively similar results. 
Our main conclusion is that the PLSM and PNJL models lead to qualitatively and to some extent quantitatively (within 10\% or less) similar phase structure at and away from the physical value of $m_\pi$. This observation strengthens the viability of these models as an effective description of the thermodynamics of two-flavor QCD at zero net-quark density. 

The paper is organized as follows: in section \ref{models} we briefly recall the basic definitions of the two models we consider and explain how the parameters are affected to take into account varying pion mass. Then in section \ref{results} we present our main results and in section \ref{checkout} our conclusions and outlook.
	  
\section{The models}
\label{models}

\subsection{Models at the physical point $m_\pi=140$ MeV}
Let us first review the general definitions of the models which we consider. We start with the description in terms of parameters fixed to correspond to the physical values reproducing the observed masses for the pion and scalar resonance $\sigma$. The PLSM model consists of the linear sigma model, a polyakov loop potential and an interaction between the two. The PNJL model is similar with the chiral part of the
Lagrangian now corresponding to the NJL model. For the derivation of the grand potential we refer to the literature, e.g. \cite{Kahara:2008yg}, and merely state the result here
\begin{eqnarray}
\Omega &=& -\frac{T\ln{\mathcal{L}}}{V} \nonumber\\
 &=& U_{\rm{chiral}}+U_\ell+\Omega_{\bar{q}q},
\end{eqnarray}
where
\begin{eqnarray}
U_{\rm{chiral}} = \frac{\lambda^2}{4}(\sigma^2+\pi^2-v^2)^2-H\sigma
\end{eqnarray}
for the PLSM model and
\begin{eqnarray}
U_{\rm{chiral}} = \frac{(m_0-M)^2}{2G}
\end{eqnarray}
for the PNJL model. 
The parameters in the above equations are fixed by the physical vacuum properties for LSM as
$H=f_\pi m_\pi^2$, where $f_\pi=93$ MeV and $m_\pi=138$ MeV. The coupling $\lambda^2 \approx 20$ is 
determined by the tree level mass $m_\sigma^2=2\lambda^2f_\pi^2+m_\pi^2$, which is set to be 600 MeV. On the other hand, for NJL part we fix 
the bare quark mass to be $m_0=5.5$ MeV and the coupling $G=10.08 $ GeV$^{-2}$. Furthermore, the constitutien mass $M$ is related to the $\bar{q}q$ condensate as $M=m_0-G\langle\bar{q}q\rangle$.

The Polyakov loop is included to both models through the mean field potential
\begin{eqnarray}
U_\ell\equiv U(\ell,\ell^\ast,T)/T^4=-\frac{b_2(T)}{2}|\ell|^2-\frac{b_3}{6}(\ell^3+
\ell^{\ast 3})+\frac{b_4}{4}(|\ell|^2)^2,
\label{polyakov_potential}
\end{eqnarray}
where
\begin{eqnarray}
b_2(T)=a_0+a_1\frac{T_0}{T}+a_2(\frac{T_0}{T})^2+a_3(\frac{T_0}{T})^3,
\end{eqnarray}
and the constants $a_i$,$b_i$ are fixed to reproduce pure gauge theory thermodynamics with 
phase transition at $T_0=270$ MeV; We adopt the values determined in \cite{Ratti}, and 
shown for completeness in table \ref{parametertable}. Here $\ell$ is the gauge invariant 
Polyakov loop in the fundamental representation, and while one could also include other loop degrees, say, adjoint or the sextet, we consider here 
only the mean field potential of the fundamental loop parametrized to describe the 
pure gauge thermodynamics. Finally there are the interactions between the Polyakov loop and chiral degrees of freedom given by the potential

\begin{eqnarray}
\Omega_{\bar{q}q} &=& 
-2N_fT\int\frac{d^3p}{(2\pi)^3}\left(\ln(1+3(\ell+\ell^\ast e^{-(E-\mu)/T})e^{-(E-\mu)/T}+e^{-3(E-\mu)/T})\right.\nonumber \\
&& \left.+\ln(1+3(\ell^\ast+\ell e^{-(E+\mu)/T})e^{-(E+\mu)/T}+e^{-3(E+\mu)/T})\right).
\label{omegaqqbar}
\end{eqnarray}

In the PNJL model the interaction potential (\ref{omegaqqbar}) has also an additional vacuum term
\begin{equation}
-6N_f\int\frac{d^3p}{(2\pi)^3}E\theta(\Lambda^2-|\vec{p}|^2) 
\end{equation}
which is controlled by the cut-off $\Lambda$. In the above equations we have $E=\sqrt{p^2+M^2}$ for the PNJL model and similarly $E=\sqrt{p^2+g^2\sigma^2}$ for the PLSM model. In the latter case the coupling constant $g$ is fixed to 3.3 corresponding to the baryon mass $\sim 1$ GeV. Thermodynamics is now determined by solving the 
equations of motion for the mean fields,
\begin{eqnarray}
\frac{\partial\Omega}{\partial\sigma}=0, ~~\frac{\partial\Omega}{\partial\ell}=0, 
~~\frac{\partial\Omega}{\partial\ell^\ast}=0,
\end{eqnarray}
and then the pressure is given by evaluating the potential on the minimum: 
$p=-\Omega(T,\mu)$. The above equations are valid at nonzero densities, but in this work we will not consider finite chemical potential but set $\mu=0$. This results in the additional simplification $\ell=\ell^\ast$.

\begin{table}
\begin{tabular}{|c | c | c | c |}
\hline
{\bf{LSM:}} & $f_\pi$ & $m_\pi$ & $m_\sigma$ \\
 & 93 MeV & 138 MeV & 600 MeV \\
 & $g$ &  $\lambda$ & $H$ \\
 & 3.3 & $\approx$ 4.44 & $\approx 1.77 \cdot 10^{-3}$ $\textrm{GeV}^3$ \\
\hline
{\bf{NJL:}} & $m_0$ & $\Lambda$ & $G$ \\
& 5.5 MeV & 651 MeV & 10.08 (GeV)$^{-2}$ \\
\hline 
{\bf{Polyakov:}} & $a_0$ & $a_1$ & $a_2$ \\
 & 6.75 & -1.95 & 2.625 \\
 & $a_3$ & $b_3$ & $b_4$ \\
 & -7.44 & 0.75 & 7.5 \\
 \hline
 \end{tabular}
\caption{The parameters used for the effective potential}
\label{parametertable}
\end{table}

\subsection{PLSM model away from the physical point}

In this work we want to explore the region in the models where the pion mass differs 
from its observed value. For this we need a consistent way of setting the model 
parameters in the non-physical region. In the PLSM model this can be achieved by finding relations
$f_\pi(m_\pi)$,  $m_\sigma(m_\pi)$ and $g(m_\pi)$ consistent with the known values at the physical point and leaving $m_\pi$ as the only tunable parameter. 
In this section we present a way to do this relying on lattice data. 

First we want to relate the pion mass $m_\pi$ and the bare quark mass $m_q$.
From \cite{Chiu:2003iw} we get the relation
\begin{equation}
m_\pi^2a^2 = A_1(m_q a)^{\frac{1}{1+\delta}} + B(m_q a)^2, 
\label{pion_mass_par}
\end{equation}
for the pion mass given in terms of the lattice spacing $a$ and
$m_q$. In (\ref{pion_mass_par}) the parameters $A_1$, $B$ and $\delta$ are
fitted from lattice data and we use the fit made in \cite{Chiu:2003iw}
\begin{equation}
\begin{aligned}
\delta &=& 0.16413 \\
A_1 &=& 0.82725 \\
B &=& 1.88687. 
\end{aligned}
\end{equation}
The lattice spacing $a$ is determined from the equation
\begin{equation}
f_\pi a = 0.06672 + 0.221820\times (m_q a)
\end{equation}
also fitted in \cite{Chiu:2003iw} to lattice data. Now to retain the relation that
$f_\pi = 93$ MeV corresponds to $m_\pi = 138$ MeV we shift the value of 
$f_\pi$ by a constant $C = 1.18$ MeV since we don't want to alter the fit
parameters or the lattice spacing obtained in \cite{Chiu:2003iw}.
This gives us the relation
\begin{equation}
m_q a = ((f_\pi + C) a - 0.06672)/0.221820
\label{m_q_a_par}
\end{equation}
with the lattice spacing $a = 0.505306$ $\textrm{GeV}^{-1}$.
Combining (\ref{pion_mass_par}) and (\ref{m_q_a_par}) gives us the pion
mass in relation with the pion decay constant. 

The sigma mass is determined through the relation
\begin{equation}
m_\sigma = \xi m_\pi^2 + D
\end{equation}
based on \cite{Kunihiro:2003yj} from which we also obtain the slope $\xi = 0.00183$ $\textrm{MeV}^{-1}$. Requiring that
$m_\pi = 138$ MeV corresponds to $m_\sigma = 600$ MeV gives us the value of the constant $D = 565.15$ MeV.

Finally, for the coupling constant $g$ we had the relation $gf_\pi = M_N/3$, where the nucleon mass $M_N$ can be
parametrized in the form
\begin{equation}
M_N = M_0 + 4C_1 m_\pi^2
\label{nuc_mass_par}
\end{equation}
as discussed in \cite{Procura:2003ig} where it has also been shown that for a good description of nucleon mass one should include terms up to and including ${\mathcal{O}}(m_\pi^4)$ which are obtained from chiral perturbation theory at ${\mathcal{O}}(p^4)$. We here truncate this fit to ${\mathcal{O}}(m_\pi^2)$, which gives a fairly poor description of $M_N(m_\pi)$ but the resulting effect on $g$ is small, at 30\% level and we will comment on the possible effect this has on our numerical results. Also, we want to keep the number of parameters to the minimum, so this approximation seems a reasonable starting point. To retain the relation ($m_\pi = 138$ MeV, $f_\pi = 93$ MeV) $\Leftrightarrow$ $g = 3.3$ we set the parameters
in (\ref{nuc_mass_par}) to $M_0 = 860$ MeV and $C_1 = 0.8$ $\textrm{GeV}^{-1}$.

\subsection{PNJL model away from the physical point}

\begin{figure}[htb]
\centering

\includegraphics[width=10cm]{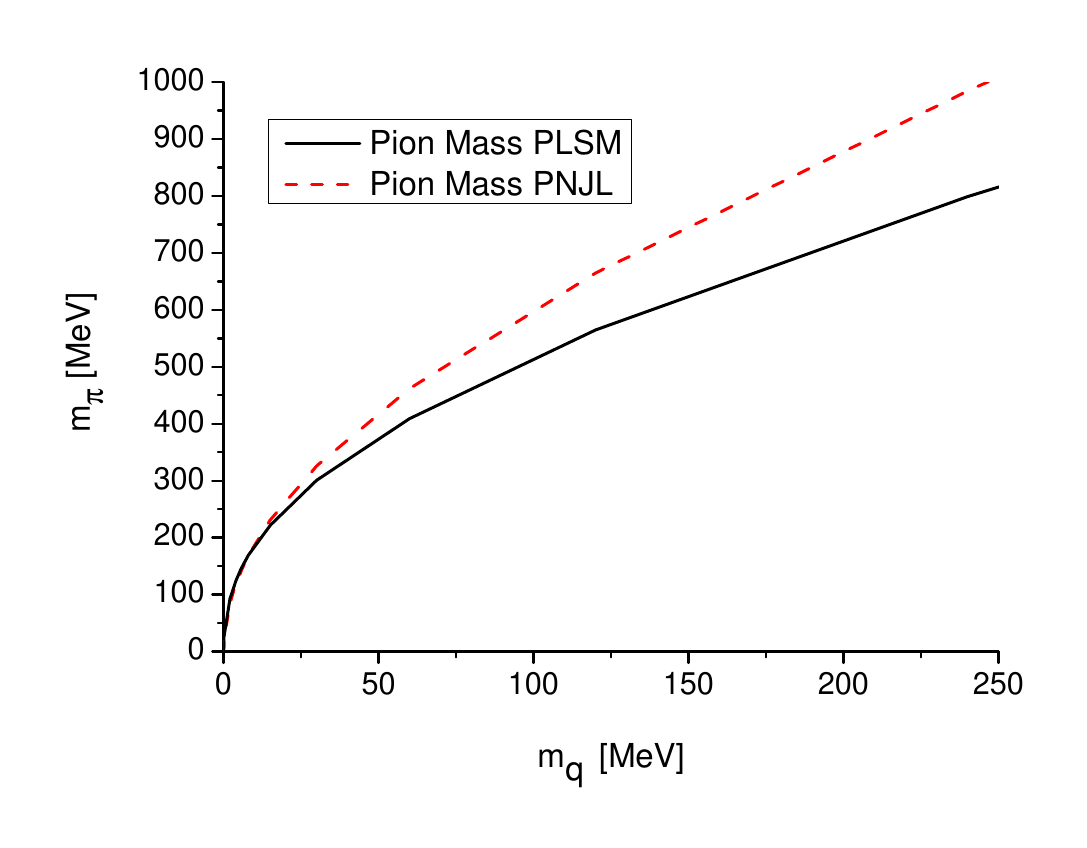}
  
\vskip-0.4truecm
\caption{Pion mass as a function of the bare quark mass in the PLSM and PNJL models.}

\label{mqvsmp}
\end{figure}

In the PLSM model we first used the lattice relation (\ref{pion_mass_par})
to connect the pion mass to the bare quark mass. In the PNJL model the bare quark mass is a direct input parameter and the
pion mass arises from the model so they are already related. The PNJL model yields a relation for the masses that is consistent
with the lattice result (\ref{pion_mass_par}). A comparison between these two
is shown in Figure \ref{mqvsmp} where the PLSM line now corresponds to
the lattice formula (\ref{pion_mass_par}). The deviation between the models is small, but grows slowly towards
higher $m_q$ being $\approx$ 10 $\%$ at $m_q = 60$ MeV and $\approx$ 20 $\%$ at $m_q = 250$ MeV.
Note that we obtain this consistency without tuning the coupling or the cutoff. Actually, we have checked numerically that as a function of both $G$ and $\Lambda$, the values in Table \ref{parametertable} give the best agreement for the pion mass with the lattice curve. 
Altering $G$ or $\Lambda$ from these values will increase $m_\pi$ for a given $m_q$ and thus adjusting these parameters would only move $m_\pi(m_q)$ further away from the corresponding lattice curve.  

Now we have a parametrization of both models in a way which allows us to study chiral symmetry breaking through one parameter, namely the quark mass $m_q$. In the PLSM model we have exchanged this parameter with the mass of the pion through a formula consistent with lattice result since in the case of PLSM model this is most convenient as the pion appears as an explicit degree of freedom. We now turn to investigation of explicit chiral symmetry breaking and its effect on the thermodynamics of these models.

\section{Effect on thermodynamics}
\label{results}

\begin{figure}[htb]
\centering
  \subfigure{
  \hskip-1.0truecm
  \includegraphics[width=8.5cm]{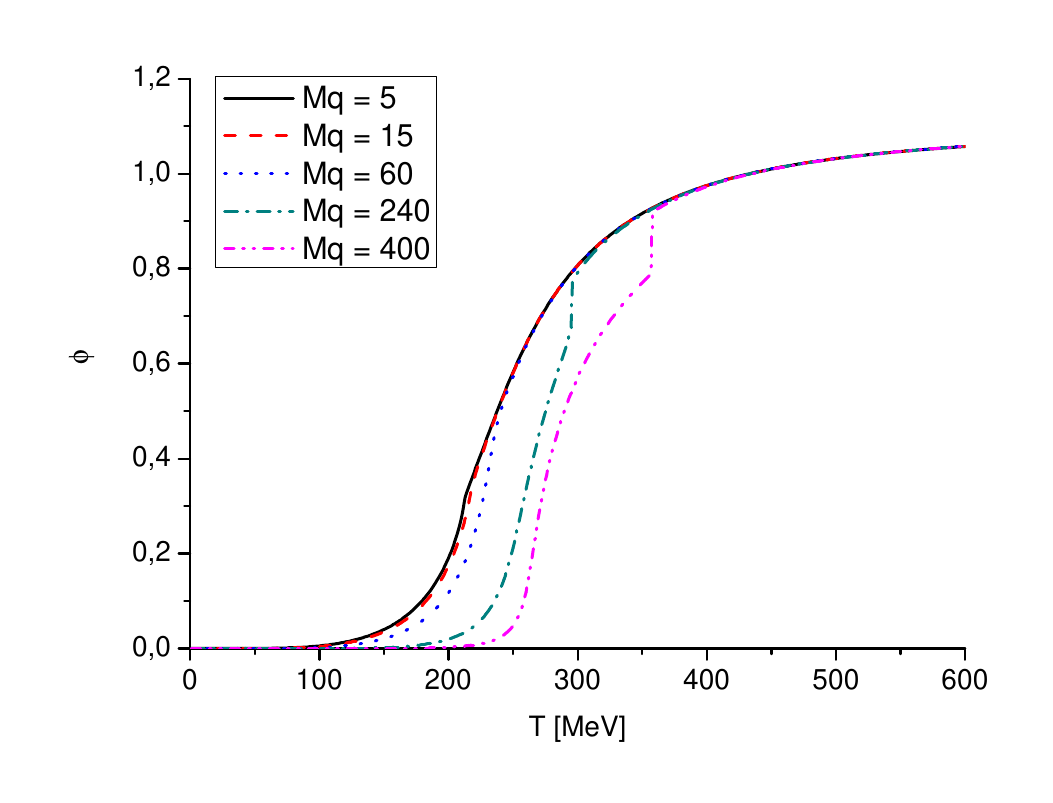}}
	\qquad
	\hskip-1.2truecm
	\subfigure{\includegraphics[width=8.5cm]{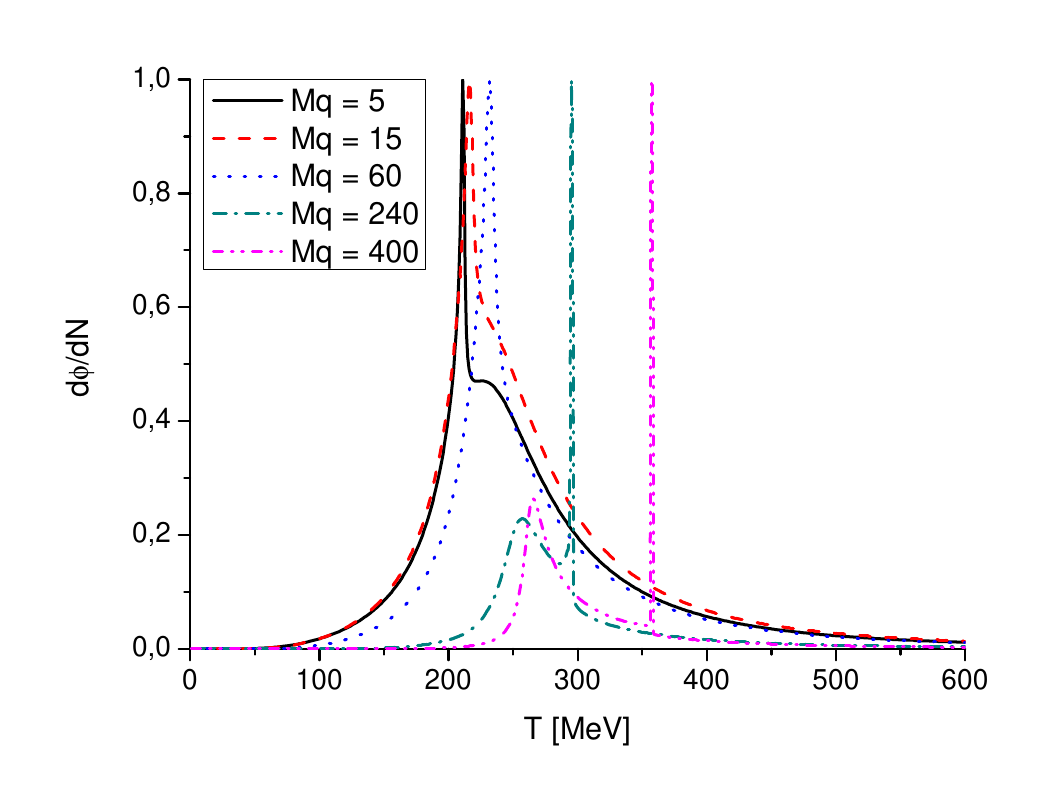}}
\vskip-0.4truecm
\caption{The Polyakov field $\phi$ and its derivative in the PLSM model for various values of $m_q$. The
derivatives have been normalized to unity for clarity.}
\label{PLSMPhi}
\end{figure}

\begin{figure}[htb]
\centering
  \subfigure{
  \hskip-1.0truecm
  \includegraphics[width=8.5cm]{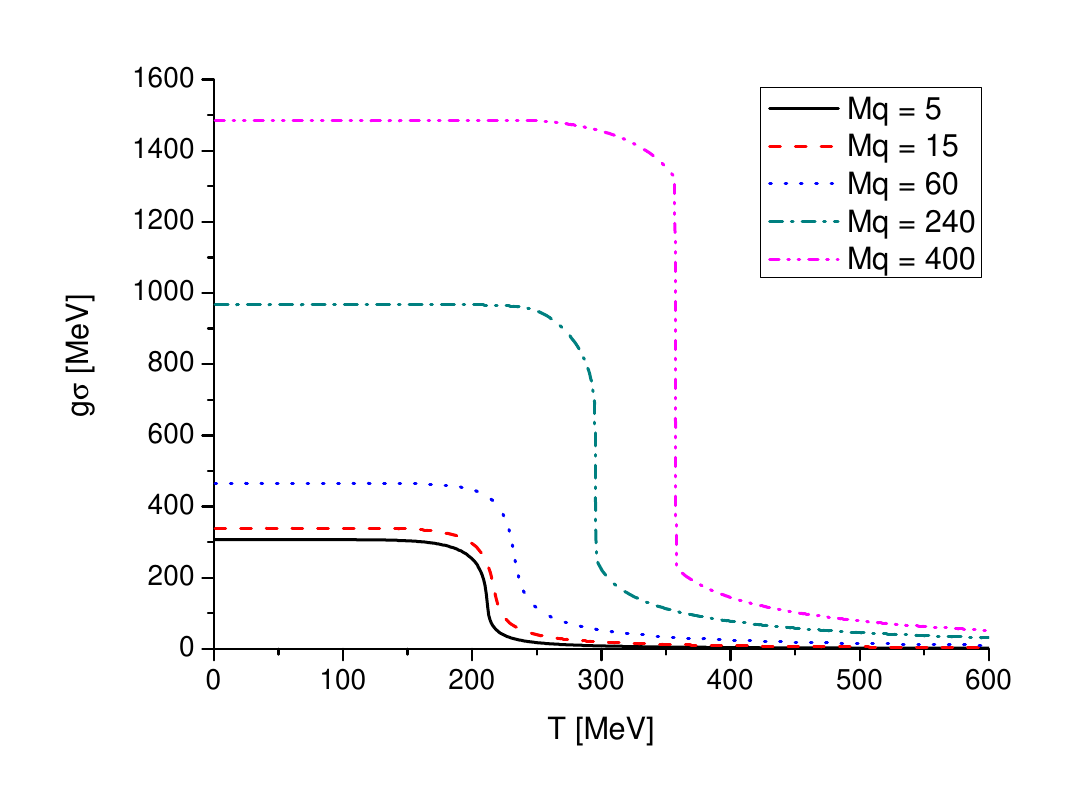}}
	\qquad
	\hskip-1.2truecm
	\subfigure{\includegraphics[width=8.5cm]{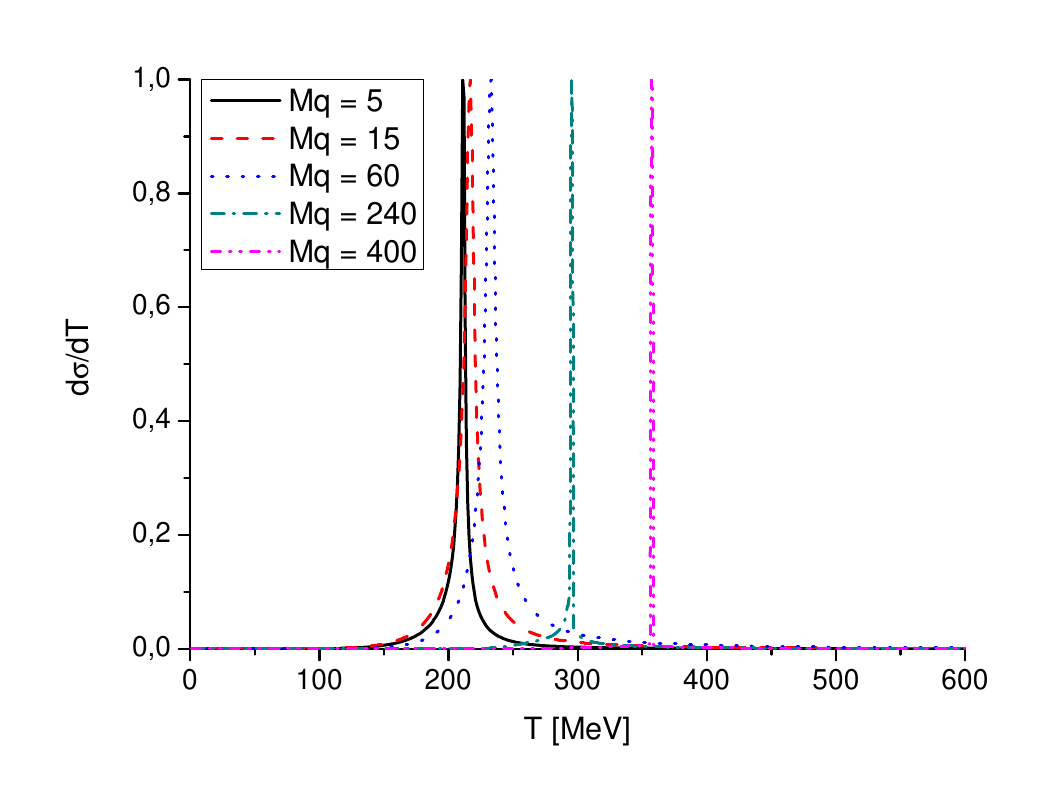}}
\vskip-0.4truecm
\caption{Constituent mass and its derivative in the PLSM model for various values of $m_q$. The
derivatives have been normalized to unity for clarity.}
\label{PLSMCMass}
\end{figure}

Using the parametrization described in the previous section we can now study the relation between the pion mass and the critical temperatures of the models. There are a priori two transitions in the models: The chiral transition due to (approximate) restoration of chiral symmetry at finite temperature and the deconfinement transition encoded into the parameters of the potential for the polyakov field $\ell$. To both of these transitions one can assign their own critical temperature. Since over most of the parameter space the transition is a crossover, the definition of a critical temperature is vague. We have taken as definition of the critical temperature of the transition to be the temperature in which the temperature derivative of the corresponding field has its peak value. This definition however has some problems since a rapid increase in the field will yield a peak in the derivative regardless of the absolute value of the field at this point. To illustrated possible ambiguities, consider Figure \ref{PLSMPhi}
where the polyakov field $\ell$ and its temperature derivative are shown. 
At large $m_q$ a double peak structure is visible in the derivative with a  broad peak and a sharp peak (a similar effect arises also when $m_q$ is less than 5 MeV, but not shown in Figure \ref{PLSMPhi}). When one looks at the values of the field 
$\phi$, the sharp peaks correspond to the discontinuous jumps in the field at large $m_q$ while the  broader peaks signify the location of the largest  continuous rise in the value of the field. It is not clear which of these
rapid changes should actually be used to determine the critical temperature. 
Some insight to the problem can be found by looking at the Figure \ref{PLSMCMass},
where the constituent mass and its derivative are shown. One immediately notices that at large $m_q$ the discontinuities in the constituent
mass in Figure \ref{PLSMCMass} correspond to the jumps in the Polyakov field $\phi$ in Figure \ref{PLSMPhi}. This suggests that the sharper
peaks in the Polyakov field derivative are merely an effect produced by the rapidly changing constituent mass and its interaction with the Polyakov
field. Also since the overall behaviour of the Polyakov field at different values of $m_q$ is not significantly altered by the appearence of these sharp
peaks it seems that the broader peak is a better indicator for the critical temperature. This can be further justified by considering that deconfinement
is characterised by $\ell\sim 1$ and the broader peak is roughly at $\ell\sim 1/2$ which is the point where the
system is becoming dominantly deconfined in contrast to being confined when $\ell\sim 0$. From here on we will call the temperature associated with the
broader peak as the (primary) $T_c$ and temperature associated with the sharper peak the induced $T_c$. We observed a similar double
peak structure also in \cite{Kahara:2008yg} and it seems to be a feature of these effective models, but may not be true in two-flavor QCD.

Note that the $T_c$ determined from the derivative of the chiral field is dependent on the bare quark mass, or equivalently on $m_\pi$. For PLSM this behavior can be affected by increasing the value of coupling $g$: larger the value of $g$, slower the increase of $T_c$ as $m_\pi$ increases. However, since $g\propto M_N$ it turns out that our simple parametrization yields larger value for $g$ than what one would get if one would fit $M_N$ more carefully from lattice \cite{Procura:2003ig}. Projecting this on our analysis implies that while we observe PLSM and PNJL transition temperatures to behave qualitatively similarly up to at least $m_\pi = 800$ MeV, it is plausible that quantitatively the critical temperature in PLSM might grow more rapidly with $m_q$ than in PNJL model. Larger values of $g$ also lead to sharper transitions and may be responsible for the discontinuous
transitions encountered in the PLSM case.

\begin{figure}[htb]
\centering
  \subfigure{
  \hskip-1.0truecm
  \includegraphics[width=8.5cm]{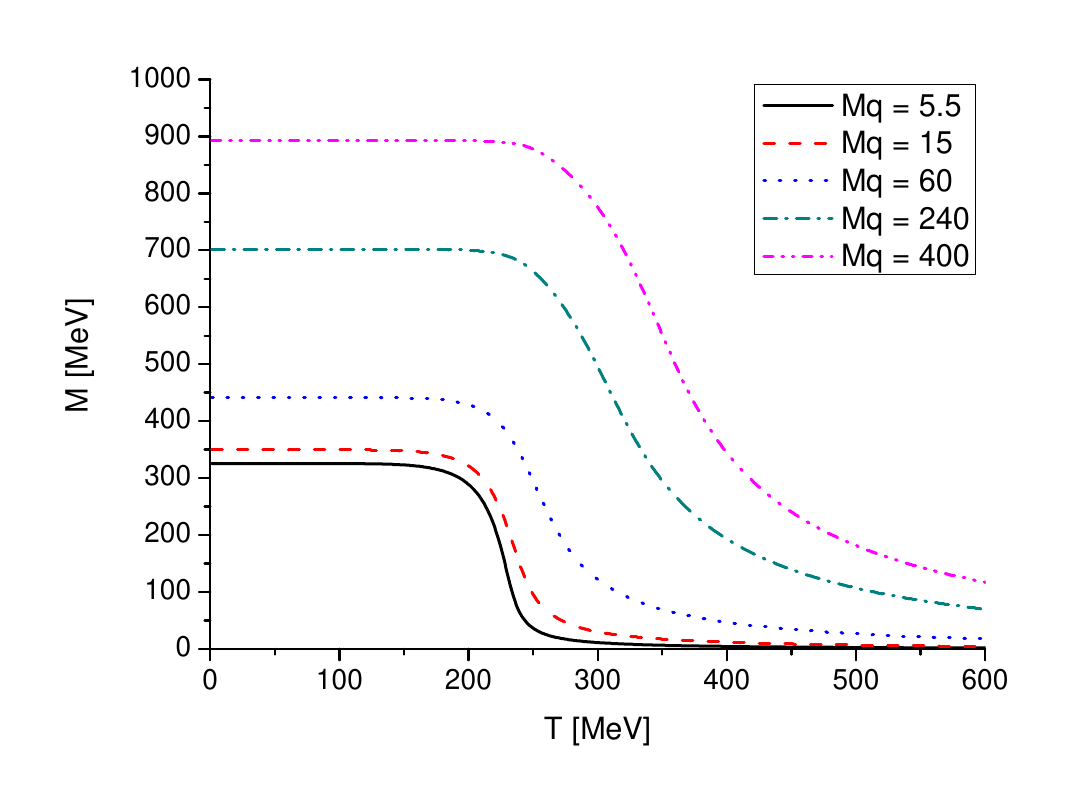}}
	\qquad
	\hskip-1.2truecm
	\subfigure{\includegraphics[width=8.5cm]{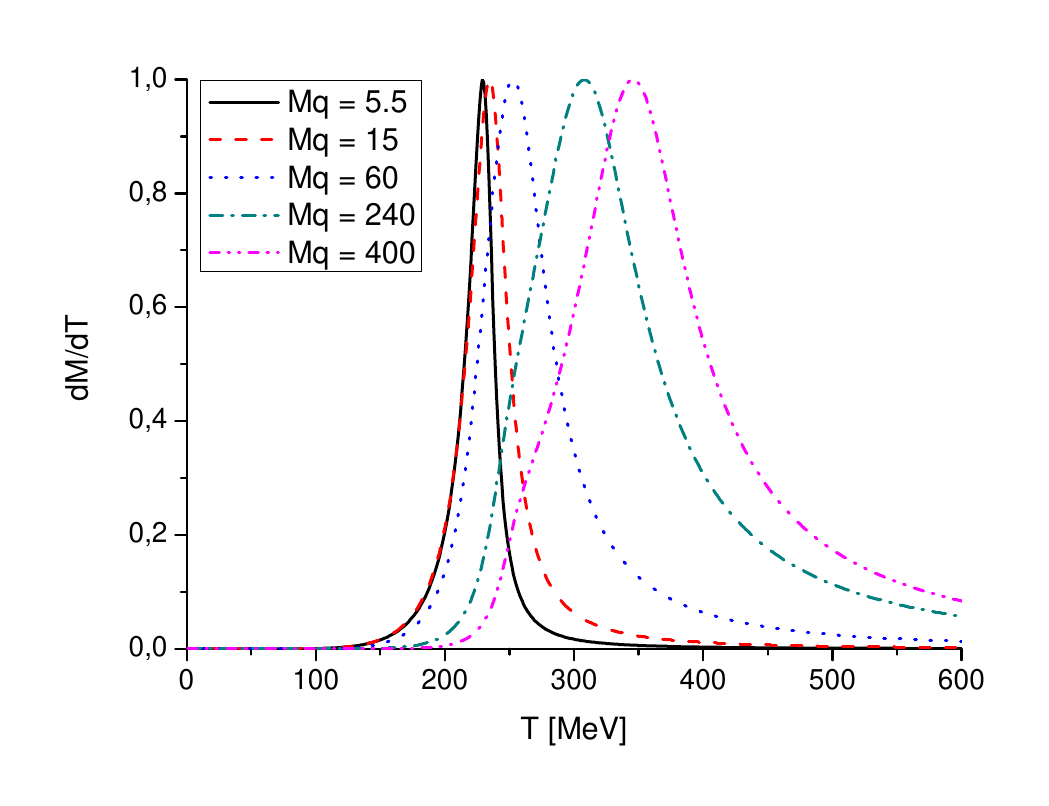}}
\vskip-0.4truecm
\caption{Constituent mass and its derivative in the PNJL model for various values of $m_q$. The
derivatives have been normalized to unity for clarity.}
\label{PNJLCMass}
\end{figure}

Turning to the PNJL case, then, we first observe that the double peak problem does not arise here, although observed also in the PNJL model
at finite $\mu$ in \cite{Kahara:2008yg}. 
A related difference with respect to the PLSM case is observed comparing Figures \ref{PLSMCMass} and \ref{PNJLCMass}: The constituent mass behaves very differently in the two models. In the PNJL model the chiral transition softens as we go towards higher $m_q$ which can be seen
as a broadening of the peaks in the derivative. In the PLSM case, on the other hand, the transition was observed to become sharper and finally discontinuous as $m_q$ is ramped up. It can be attributed to this difference that the double peak structure in the Polyakov field arises in the PLSM model but not in the PNJL model. Despite the difference in the sharpness of the chiral transition the transition temperatures determined from the
peaks of the derivatives are relatively close to each other.

\begin{figure}[htb]
\centering
  \subfigure{
  \hskip-1.0truecm
  \includegraphics[width=8.5cm]{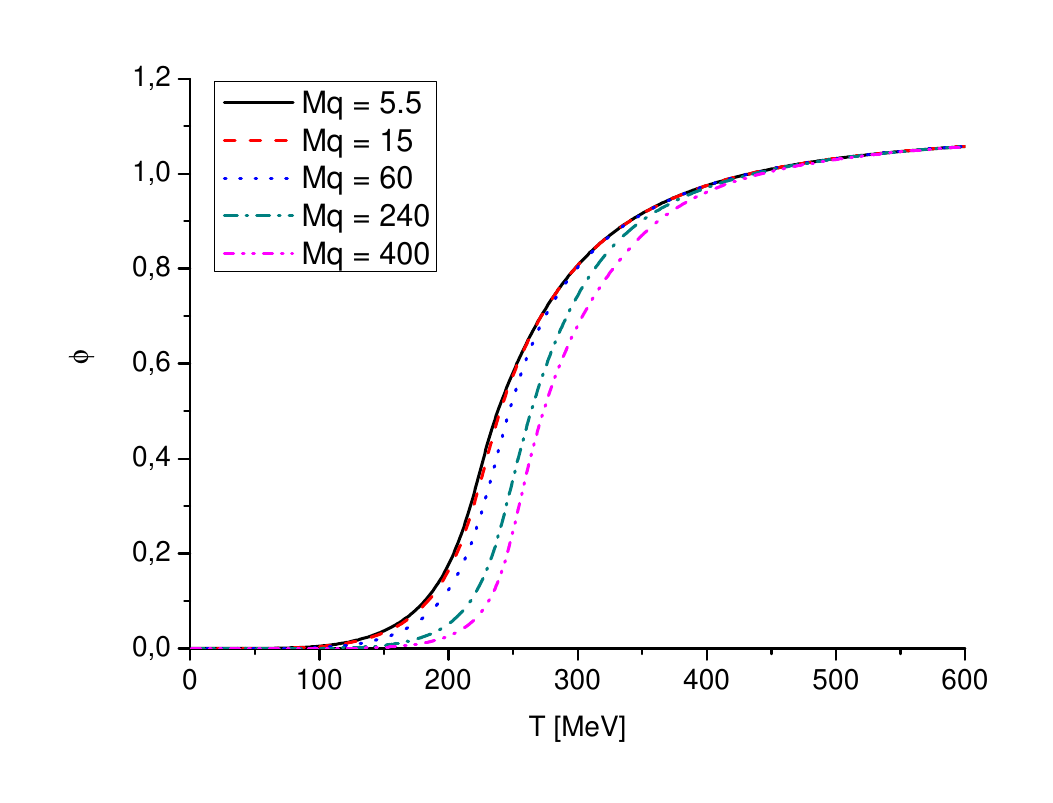}}
	\qquad
	\hskip-1.2truecm
	\subfigure{\includegraphics[width=8.5cm]{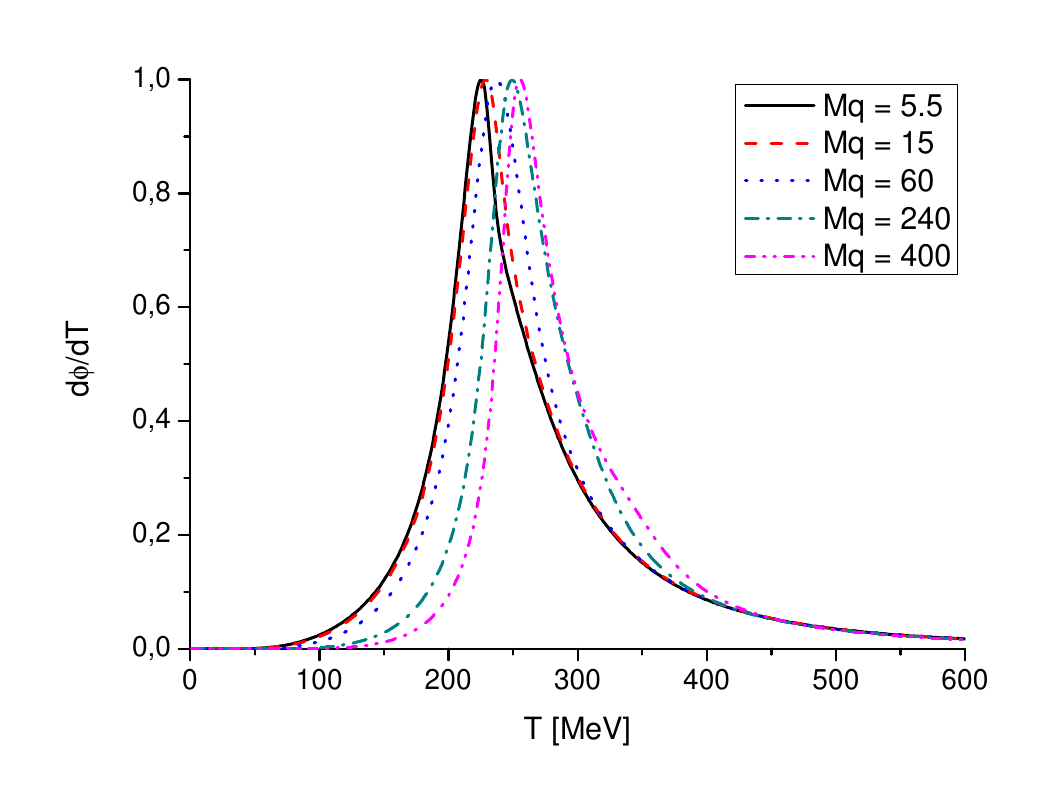}}
\vskip-0.4truecm
\caption{The polyakov field $\phi$ and its derivative in the PNJL model for various values of $m_q$. The
derivatives have been normalized to unity for clarity.}
\label{PNJLPhi}
\end{figure}  

\begin{figure}[htb]
\centering
  \subfigure{
  \hskip-1.0truecm
  \includegraphics[width=8.5cm]{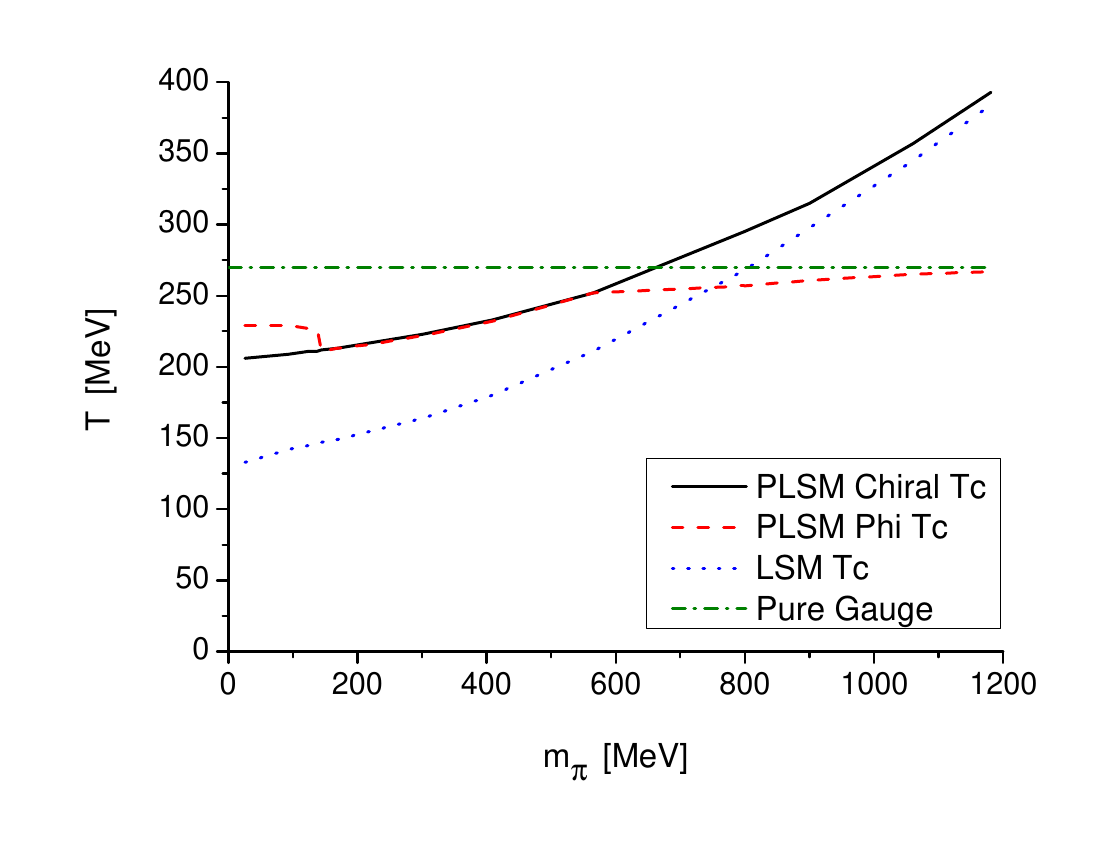}}
	\qquad
	\hskip-1.2truecm
	\subfigure{\includegraphics[width=8.5cm]{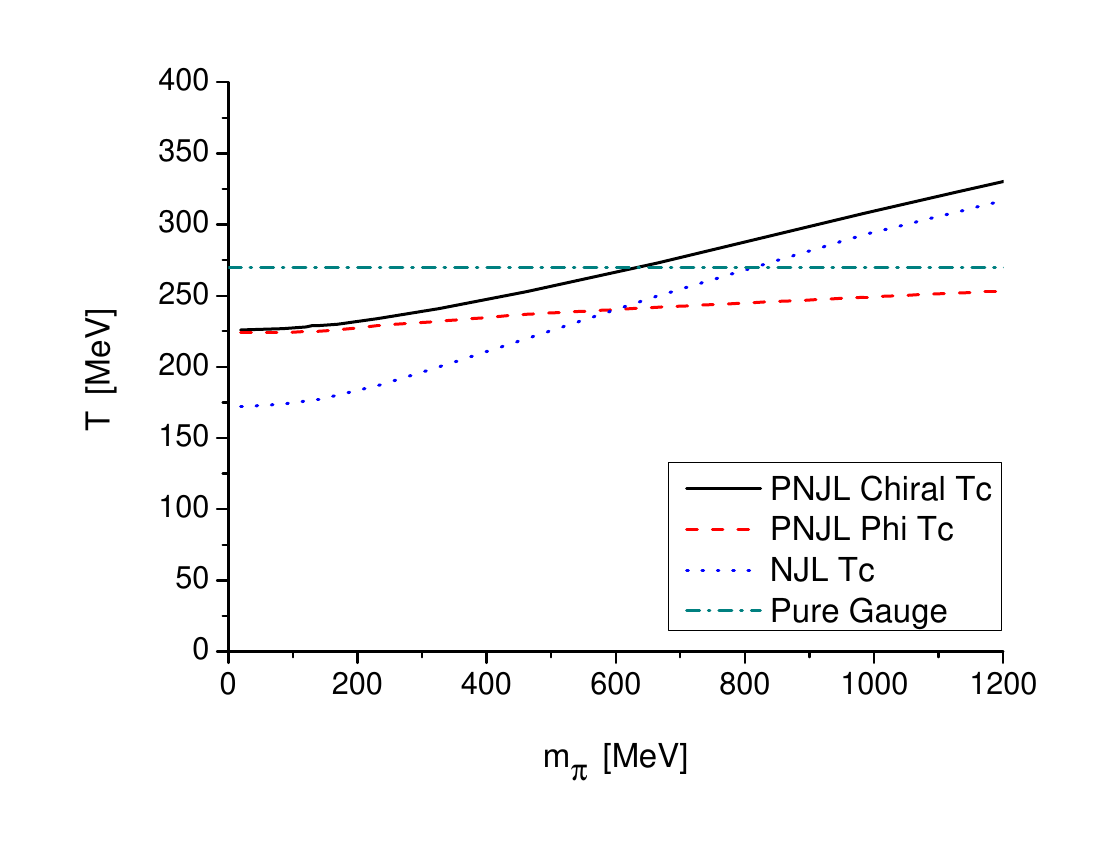}}
\vskip-0.4truecm
\caption{The critical temperatures of the chiral and the
deconfinement transitions shown against the pion mass in both models.}
\label{transitions}
\end{figure}

Let us then sketch the phase diagrams of these theories. The phase boundaries of both models are shown in Figure \ref{transitions} for a range of pion masses. From previous studies we know that when $m_\pi = 138$ MeV both, chiral and polyakov, transitions occur at a temperature of 211 MeV in the PLSM model and at 229 MeV in the PNJL model. Away from the physical point these two models behave similarly as a function of pion mass to about $m_\pi = 800$ MeV from which upwards the PLSM chiral $T_c$ has a rising curve while the PNJL model curve starts to level. Also the Polyakov field behaviour is similar in both models, although it seems to be more strongly coupled to the chiral sector in the PLSM case when $m_\pi < 600$ MeV. At larger $m_\pi$ the Polyakov field 'decouples' from the chiral sector in both cases and its $T_c$ seems to level off at near the pure gauge value. Note
that in Figure \ref{transitions} for the field $\phi$ in the PLSM, the critical temperature is the primary $T_c$ discussed above. The induced $T_c$ follows the chiral transition curve and is not shown separately.
In addition to the PLSM and PNJL models Figure \ref{transitions} shows the critical temperatures for the bare NJL and LSM models as well as for the pure gauge potential. Comparing the NJL and pure gauge critical temperatures with the PNJL model, one can see that the critical temperatures are shifted closer together at small $m_\pi$.
One can also see that the interaction induces a $m_\pi$ dependence on the Polyakov field critical temperature which is independent of $m_\pi$ in the pure gauge case. Also naturally the $m_\pi$ dependence of the chiral field is altered. At large $m_\pi$ both of these
effects die out as the different transitions separate. The chiral part of PNJL approaches the NJL curve and the Polyakov field draws near to the pure
gauge curve. The same things can be said from the PLSM case where the chiral and Polyakov sector critical temperatures are shifted towards each other at small $m_\pi$, but at larger $m_\pi$ separate and approach the bare LSM and pure gauge curves respectively. Only difference being again
the stronger interaction between the two sectors in the range $200-600$ MeV of the pion mass.

\section{Conclusions}
\label{checkout}

We have considered the PLSM and PNJL models at non-physical pion masses in order to determine how these models react to explicit chiral symmetry breaking. The motivations for this study were in general terms to determine how well these models encode the QCD dynamics of chiral symmetry breaking and, in more specific terms, to test if the intertwining of deconfinement and chiral symmetry restoration in these models survives away from the physical value of the pion mass. For the PLSM model this analysis was made possible by a lattice-based parametrization which allowed us to set the model parameters in a consistent way away from the physical pion mass. In the PNJL model the pion mass arises from the model and can be changed by adjusting the bare quark mass which is a direct input parameter of the model. As we have shown, both models lead to similar behavior for $m_\pi (m_q)$ and are in this extent consistent with existing lattice data.

In section \ref{results} we studied the thermodynamics, or more precisely the transition temperatures and phase diagrams, of the two
models as a function of pion mass. We found that the behaviour of the polyakov sector is almost identical in the two models, as expected. The only real difference being the notable interaction between the chiral and polyakov sectors at low pion masses in the PLSM model. However the overall behaviour of the polyakov field was not largely affected by this interaction and the primary Polyakov $T_c$ agreed with the one acquired from the PNJL model. 
In contrast to the Polyakov sector, the chiral transition temperatures showed a different behaviour in the two models at pion masses larger than 800 MeV, but this should rather be taken as indication of the fact that such large pion masses are beyond the validity of simple chiral models. 

\begin{figure}[htb]
\centering
  \includegraphics[width=10cm]{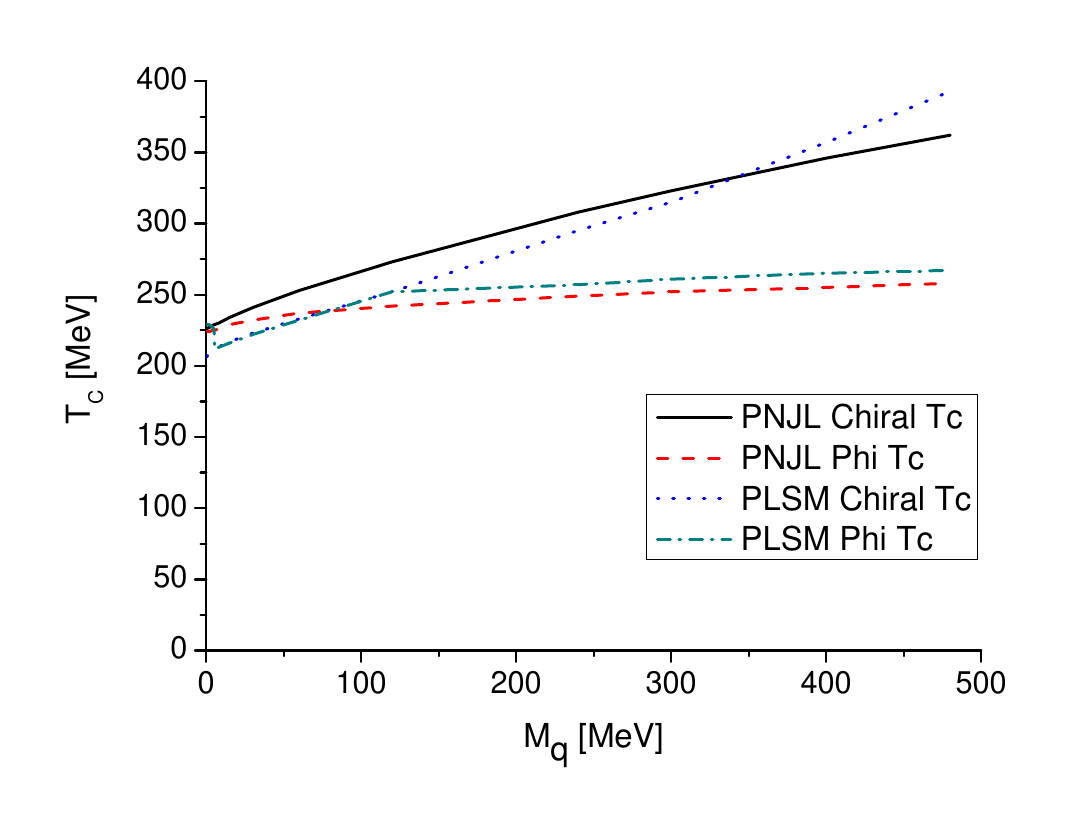}
\caption{The critical temperatures of the chiral and the
deconfinement transitions shown against the quark mass in both models.}
\label{Tc_mq}
\end{figure}

For the pion masses above 400 MeV our analysis for the PLSM model confirms the results found in \cite{Dumitru:2003cf} and qualitatively the results obtained for the PNJL model are similar. Since such masses make the applicability of these chiral models questionable, one should concentrate more on the range below $m_\pi=400$ MeV. In this range we observe that in qualitative agreement both PLSM and PNJL models imply coincident chiral symmetry breaking and deconfinement as well as modestly increasing critical temperature as a function of the pion mass. Within this range the models agree quantitatively within $\sim 10\%$ although it should be noted that the critical temperature grows more rapidly with increasing pion mass in PLSM model than in PNJL model. Since the quark mass can be thought of a more basic parameter of these models than the pion mass and since the pion mass corresponding to given quark mass differ in these two models, it is informative to also plot $T_c$ as a function of the quark mass. This is shown in Figure \ref{Tc_mq}, and clearly confirms the result that near the chiral limit the crossover transition $T_c$ can be determined either using the chiral or Polyakov loop degrees of freedom. As one moves further away from the chiral limit, above $m_q\sim 100$ MeV corresponding to $m_\pi\sim 400$ MeV (see Figure \ref{mqvsmp}), the description using chiral fields becomes more uncertain. The results obtained from the Polyakov loop for $T_c$ are similar in the two models which is not surprising since the mean field potential for the Polyakov loop is the same in both models and only weakly affected by the value of the quark mass through the interactions with chiral fields.

Existing lattice data for two-flavor QCD thermodynamics at pion masses above 400 MeV \cite{Karsch:2000kv} together with our results in this mass range suggests that the dependence of the critical temperature on the pion mass is better described by the Polyakov loop dynamics. In the model framework considered here this is natural consequence of the underlying coupled dynamics. Lattice data closer towards the physical value of the pion mass would be highly desirable in order to better constrain these effective models.

\acknowledgments
The financial support for T.K. from the V\"ais\"al\"a foundation is gratefully acknowledged.

\end{document}